
\input harvmac
\input amssym.def
\input amssym
\baselineskip 14pt
\magnification\magstep1
\parskip 6pt
\newdimen\itemindent \itemindent=32pt
\def\textindent#1{\parindent=\itemindent\let\par=\resetpar%
\indent\llap{#1\enspace}\ignorespaces}

\let\oldpar=\par
\def\resetpar{\oldpar\parindent=20pt\let\par=\oldpar}

\font\ninerm=cmr9 \font\ninesy=cmsy9
\font\eightrm=cmr8 \font\sixrm=cmr6
\font\eighti=cmmi8 \font\sixi=cmmi6
\font\eightsy=cmsy8 \font\sixsy=cmsy6
\font\eightbf=cmbx8 \font\sixbf=cmbx6
\font\eightit=cmti8
\def\eightpoint{\def\rm{\fam0\eightrm}
  \textfont0=\eightrm \scriptfont0=\sixrm \scriptscriptfont0=\fiverm
  \textfont1=\eighti  \scriptfont1=\sixi  \scriptscriptfont1=\fivei
  \textfont2=\eightsy \scriptfont2=\sixsy \scriptscriptfont2=\fivesy
  \textfont3=\tenex   \scriptfont3=\tenex \scriptscriptfont3=\tenex
  \textfont\itfam=\eightit  \def\it{\fam\itfam\eightit}%
  \textfont\bffam=\eightbf  \scriptfont\bffam=\sixbf
  \scriptscriptfont\bffam=\fivebf  \def\bf{\fam\bffam\eightbf}%
  \normalbaselineskip=9pt
  \setbox\strutbox=\hbox{\vrule height7pt depth2pt width0pt}%
  \let\big=\eightbig  \normalbaselines\rm}
\catcode`@=11 %
\def\eightbig#1{{\hbox{$\textfont0=\ninerm\textfont2=\ninesy
  \left#1\vbox to6.5pt{}\right.\n@@space$}}}
\def\vfootnote#1{\insert\footins\bgroup\eightpoint
  \interlinepenalty=\interfootnotelinepenalty
  \splittopskip=\ht\strutbox %
  \splitmaxdepth=\dp\strutbox %
  \leftskip=0pt \rightskip=0pt \spaceskip=0pt \xspaceskip=0pt
  \textindent{#1}\footstrut\futurelet\next\fo@t}
\catcode`@=12 %
\def \d{{\rm d}}

\def \si{\sigma}

\def \pr{\partial}
\def \d{{\rm d}}

\def \hs{\hat s}
\def \hr{\hat r}
\def \br{\bar r}

\def \bl{{\lambda}}

\def \l{\big \langle}
\def \r{\big \rangle}

\def \vep{\varepsilon}
\def \half{{\textstyle {1 \over 2}}}

\def \ts{\textstyle}

\def \F{{\cal F}}

\def \N{{\cal N}}

\def\limu#1{\mathrel{\mathop{\sim}\limits_{\scriptscriptstyle{#1}}}}

\font \bigbf=cmbx10 scaled \magstep1

\lref\LR{K. Lang and W. R\"uhl,  The Critical O(N) Sigma Model at Dimensions
2$<$D$<$4: Fusion Coefficients and Anomalous Dimensions,
Nucl. Phys. {B400} (1993) 597.}
\lref\Fone{S. Ferrara, A.F. Grillo, R. Gatto and G. Parisi, Covariant Expansion
of the Conformal Four-Point Function, Nucl. Phys. B49 (1972) 77\semi
S. Ferrara, A.F. Grillo, R. Gatto and G. Parisi, Analyticity Properties and
Asymptotic Expansions of Conformal Covariant Green's Functions, Nuovo Cimento 
19A (1974) 667\semi
S. Ferrara, R. Gatto and A.F. Grillo, Properties of Partial-Wave Amplitudes
in Conformal Invariant Theories, Nuovo Cimento 26A (1975) 226.}
\lref\Ftwo{S. Ferrara, A.F. Grillo and R. Gatto, Ann. Phys. 76 (1973) 161.}
\lref\Dob{V.K. Dobrev, V.B. Petkova, S.G. Petrova and I.T. Todorov,
Dynamical Derivation of Vacuum Operator Product Expansion in Conformal Field
Theory, Phys. Rev. D13 (1976) 887.}
\lref\one{F.A. Dolan and H. Osborn, Implications of $\N=1$ Superconformal
Symmetry for Chiral Fields, Nucl. Phys. B593 (2001) 599, hep-th/0006098.}
\lref\Grad{{\it Tables of Integrals, Series, and Products}, 5th ed., 
I.S. Gradshteyn and I.M. Ryzhik, ed. A. Jeffrey, 
(Academic Press, San Diego, 1994).}
\lref\DO{F.A. Dolan and H. Osborn, Conformal Four Point Functions and the
Operator Product Expansion, Nucl. Phys. B599 (2001) 459,  hep-th/0011040.}
\lref\SCFT{F.A. Dolan and H. Osborn, Superconformal Symmetry, Correlation 
Functions and the Operator Product Expansion, Nucl. Phys. B629 (2002) 3,
hep-th/0112251.}
\lref\Arut{G. Arutyunov and E. Sokatchev, Implications of Superconformal
Symmetry for Interacting (2,0) Tensor Multiplets, Nucl.Phys. B635 (2002) 3,
hep-th/0201145.} 
\lref\Dirac{P.A.M. Dirac, Wave Equations in Conformal Space,
Annals of Math. 37 (1936) 823; {\it in} ``The
Collected Works of P.A.M. Dirac 1924-1948'', edited by R.H. Dalitz, CUP
(Cambridge) 1995.}
\lref\FGG{S. Ferrara, R. Gatto and A.F. Grillo, ``Conformal Algebra in 
Space-time and Operator Product Expansion", 
(Springer Tracts in Modern Physics, vol. 67) Springer (Heidelberg) 1973.}
\lref\Vilenkin{N.Ja. Vilenkin and A.U. Klimyk, ``Representation of Lie 
groups and special functions: recent advances'',
Kluwer Academic Publishers (Boston) 1995.}
\lref\Vretare{L. Vretare, Formulas for Elementary Spherical Functions and
Generalized Jacobi Polynomials, Siam J. on Mathematical Analysis, 15 (1984)
805.}
\lref\Howe{P.J. Heslop and P.S. Howe, Four-point functions in $N=4$ SYM,
JHEP 0301 (2003) 043, hep-th/0211252.}
\lref\Muir{R.J. Muirhead, Systems of Partial Differential Equations for
Hypergeometric Functions of Matrix Argument, The Annals of Mathematical
Statistics, 41 (1970) 991.}
\lref\Jack{H. Jack, A class of symmetric polynomials with a parameter,
Proc. Roy. Soc. Edinburgh, 69 (1970) 1.}
\lref\nirschl{M. Nirschl and H. Osborn, in preparation.}
\lref\Koo{T. Koornwinder and I. Sprinkhuizen-Kuyper, Generalized Power
Series Expansions for a Class of Orthogonal Polynomials in Two Variables,
Siam J. on Mathematical Analysis, 9 (1978) 457.}
\lref\Spec{G.E. Andrews, R.Askey and R. Roy, ``Special Functions", Cambridge
University Press (Cambridge) 1999.}
\lref\SJack{P. Desrosiers, L. Lapointe and P. Mathieu, Jack Superpolynomials,
Superpartition Ordering and Determinantal Formulas, Comm. Math. Phys. 233
(2003) 383, hep-th/0105107, Jack Polynomials in Superspace, Comm. Math.
Phys., to be published, hep-th/0209074.}

{\nopagenumbers
\rightline{DAMTP/03-91}
\rightline{hep-th/0309180}
\vskip 1.5truecm
\centerline {\bigbf Conformal Partial Waves and the Operator Product
Expansion}
\vskip  6pt
\vskip 2.0 true cm
\centerline {F.A. Dolan and H. Osborn${}^\dagger$}

\vskip 12pt
\centerline {\ Department of Applied Mathematics and Theoretical Physics,}
\centerline {Wilberforce Road, Cambridge CB3 0WA, England}
\vskip 1.5 true cm

{\eightpoint
\parindent 1.5cm{

{\narrower\smallskip\parindent 0pt
By solving the two variable differential equations which arise from
finding the eigenfunctions for the Casimir operator for $O(d,2)$ 
succinct expressions are found for the functions, conformal partial waves,
representing the contribution of an operator of arbitrary scale dimension 
$\Delta$ and spin $\ell$ together with its descendants to conformal four 
point functions for $d=4$, recovering old results, and also for $d=6$.
The results are expressed in terms of ordinary hypergeometric functions of 
variables $x,z$ which are simply related to the usual conformal
invariants. An expression for the conformal partial wave amplitude valid
for any dimension is also found  in terms of a sum over two variable 
symmetric Jack polynomials which is used to derive relations for the
conformal partial waves.



\narrower}}

\vfill
\line{${}^\dagger$ 
address for correspondence: Trinity College, Cambridge, CB2 1TQ, England\hfill}
\line{\hskip0.2cm emails:
{{\tt fad20@damtp.cam.ac.uk} and \tt ho@damtp.cam.ac.uk}\hfill}
}

\eject}
\pageno=1

\newsec{Introduction}

The operator product expansion is a crucial part of the theoretical
structure of conformal field theories. Using the operator product expansion
to analyse conformal correlation functions determines the spectrum of
operators, their scale dimensions and spins, which are present in the theory.
In essence the procedure is analogous to the the partial wave expansion
of scattering amplitudes in terms of Legendre polynomials and their
generalisations which applied to experimental data reveals resonant states
of various masses and spins. There are two critical differences in the
application to conformal amplitudes, firstly that the symmetry group is non
compact so that the expansion functions are not finite order polynomials and
secondly that for four point functions we have to deal with functions
of two variables, corresponding to the two conformal invariants in this
case.

Results \refs{\Fone,\Dob,\LR} obtained for the contribution of a single 
conformal primary operator and its 
descendants to the four point correlation function of just scalar fields
have been rather complicated, involving in general multiple series
and integral representations. Nevertheless we recently found a quite
simple formula for the result for conformal theories in four dimensions
which just involves ordinary hypergeometric functions {\DO}. These
expressions have been used to analyse the correlation functions for
$\N=4$ superconformal chiral primary operators and reveal anomalous
dimensions for arbitrary spin both perturbatively and in ther large $N$
limit using results from the AdS/CFT correspondence \SCFT.

Nevertheless conformal fields theories are potentially of interest in 3,5
and also 6 dimensions. In the last case there are isolated non trivial field
theories with $(2,0)$  superconformal symmetry which have no free
field limit but for which large $N$ results have been obtained, \Arut. The
results obtained in \DO\ depend on using new variables $x,z$ which are
simply related to the usual conformal invariants but the method used
depends on solving a non trivial recurrence relation which is not
easy to generalise. Here we adopt an alternative approach by considering
the equations obtained by looking for the eigenfunctions, with appropriate
boundary conditions, of the $d$-dimensional conformal group $O(d,2)$ Casimir
operator.

As has been known for a long time \refs{\Dirac,\FGG} it is convenient to use 
homogeneous coordinates $\eta^A$ on the projective null cone 
$\eta^2 = g_{AB}\eta^A \eta^B =0$
with $\eta^A \sim \lambda \eta^A$ and the $d{+2}$-dimensional metric
$g_{AB} = {\rm diag.}(-1,1,\dots,1,-1)$. Conformal transformations act on 
$\eta$ linearly and the corresponding generators are
\eqn\gen{
L_{AB} = \eta_A {\pr \over \pr \eta^B} - \eta_B {\pr \over \pr \eta^A} \, .
}
Quantum fields are then extended to homogeneous functions of $\eta$, for
a scalar field $\phi$ of scale dimension $\Delta$, $\phi(\lambda \eta) = 
\lambda^{-\Delta} \phi(\eta)$.

The four point correlation function for
such scalar fields may then in general be expressed as
\eqn\fourp{\eqalign{
\l & \phi_1(\eta_1) \, \phi_2(\eta_2) \,\phi_3(\eta_3) \,\phi_4(\eta_4) \r \cr
&{} = {1\over (\eta_1{\cdot \eta_2})^{{1\over 2}(\Delta_1 + \Delta_2)}
\, (\eta_3{\cdot \eta_4})^{{1\over 2}(\Delta_3 + \Delta_4)}} \,
\bigg ( {\eta_2{\cdot \eta_4} \over \eta_1{\cdot \eta_4}}
\bigg)^{\!{1\over 2}\Delta_{12}}
\bigg ( {\eta_1{\cdot \eta_4} \over \eta_1{\cdot \eta_3}}
\bigg)^{\!{1\over 2}\Delta_{34}} F(u,v) \, , \cr}
}
where
\eqn\Ddiff{
\Delta_{ij} = \Delta_i - \Delta_j \, ,
}
and $u,v$ are two conformal invariants
\eqn\defuv{
u= {\eta_1{\cdot \eta_2} \, \eta_3{\cdot \eta_4} \over
\eta_1{\cdot \eta_3} \, \eta_2{\cdot \eta_4} } \, , \qquad
v= {\eta_1{\cdot \eta_4} \, \eta_2{\cdot \eta_3} \over
\eta_1{\cdot \eta_3} \, \eta_2{\cdot \eta_4} } \, .
}
For the contribution of a single operator  of scale dimension $\Delta$ and
spin $\ell$ in the operator product expansion of $\phi_1(\eta_1) \, 
\phi_2(\eta_2)$ to the four point function, \fourp\ then requires a function
$F=G_\Delta^{(\ell)}$ so that
\eqn\eigen{\eqalign{
L^2 \l \phi_1(\eta_1) \, \phi_2(\eta_2) \,\phi_3(\eta_3) \,\phi_4(\eta_4) \r
= {}& - C_{\Delta,\ell} \l  \phi_1(\eta_1) \, \phi_2(\eta_2) \,
\phi_3(\eta_3) \,\phi_4(\eta_4) \r \, , \cr
C_{\Delta,\ell} = {}& \Delta ( \Delta -d ) + \ell( \ell + d-2) \, , \cr}
}
where
\eqn\LL{
L^2 = \half L^{AB} L_{AB} \, , \qquad L_{AB} = L_{1AB} + L_{2 AB} \, ,
}
for $L_{iAB}$ the generator formed from $\eta_i$ as in \gen\ (conformal
invariance of course requires $\sum_i L_{iAB}
\l \phi_1(\eta_1) \, \phi_2(\eta_2) \,\phi_3(\eta_3) \,\phi_4(\eta_4) \r =0$).
$G_\Delta^{(\ell)}(u,v)$ are here referred to as conformal partial waves
and \eigen\ gives a corresponding eigenvalue equation for them,
\eqnn\eigenF
$$\eqalignno{
& L^2 G_\Delta^{(\ell)} 
- (\Delta_{12}-\Delta_{34}) \bigg ( (1+u-v) \Big ( u {\pr \over \pr u}
+  v {\pr \over \pr v} \Big ) - (1 - u -v) {\pr \over \pr v} \bigg )
G_\Delta^{(\ell)} \cr
& \quad {} - \half \Delta_{12} \Delta_{34} (1+u-v)G_\Delta^{(\ell)} 
= -  C_{\Delta,\ell} G_\Delta^{(\ell)} \, , & \eigenF \cr}
$$
where for any $F(u,v)$,
\eqn\LF{\eqalign{
\half L^2 F = {}& -\big ( (1-v)^2 - u(1+v) \big ) {\pr \over \pr v}  
v {\pr \over \pr v} F - (1-u+v)  u {\pr \over \pr u}  u {\pr \over \pr u} F \cr
& {}+ 2 (1+u-v) u v{\pr^2 \over \pr u \pr v} F 
+ d \, u {\pr \over \pr u} F \, .  \cr}
}
For $u\to 0$, $G_\Delta^{(\ell)}$ is independent of $d$ 
and, with a convenient normalisation \refs{\Fone,\DO}, it has the form
\eqn\Glim{
G_\Delta^{(\ell)}(u,v) \sim u^{{1\over 2}(\Delta -\ell)} 
\big ( {- \half} (1-v) \big )^\ell
F ( \half (\Delta + \ell - \Delta_{12}), \half (\Delta + \ell + \Delta_{34});
\Delta + \ell; 1-v ) \, ,
}
where $F$ is  a hypergeometric function. The leading behaviour as $u\to 0$
and then $v\to 1$ is thus simply determined by $\Delta,\ell$. For $\ell=0$
explicit solutions as double power series in $u$ and $1-v$ are known, see \DO,
but there is no apparent concise generalisation for $\ell>0$.

The aim here is then to solve \eigenF\ in as simple a form as possible.
In the next section we show how this is possible in four and six
dimensions as products of ordinary hypergeometric functions while
in the following section for general dimension we obtain an expansion
in terms of symmetric Jack polynomials. This is used to obtain various 
relations for conformal partial waves valid in any dimension.

\newsec{Solutions in Four and Six Dimensions}

It is now apparent that finding expressions for conformal partial
waves valid for any $\ell$ is assisted by introducing, as in \DO, new 
variables $x,z$, such that
\eqn\defxz{
u = xz \, , \qquad v = (1-x)(1-z) \, .
}
These may be interpretated \Howe\ as the eigenvalues of a $2\times 2$
spinor matrix formed from $x_1,x_2,x_3,x_4$.
Manifestly all results must be symmetric under $x\leftrightarrow z$. The
eigenvalue equation \eigenF\ then becomes
\eqn\Dop{
2 D_\vep G_\Delta^{(\ell)} = C_{\Delta,\ell} G_\Delta^{(\ell)} \, ,
}
where $D_\vep$ is a symmetric differential operator
\eqnn\defD
$$\eqalignno{
D_\vep = {}& x^2 (1-x) {\pr^2 \over \pr x^2} +  z^2 (1-z) {\pr^2 \over \pr z^2}
+ c \Big ( x {\pr \over \pr x} + z {\pr \over \pr z} \Big )
- (a+b+1) \Big ( x^2 {\pr \over \pr x} + z^2 {\pr \over \pr z} \Big ) \cr
&{} - ab (x+z)
+ \vep \, {xz \over x-z} \Big ( (1-x) {\pr \over \pr x} - (1- z) 
{\pr \over \pr z} \Big ) \, , & \defD \cr}
$$
depending on parameters $a,b,c,\vep$ which for application
in \Dop\ should take the values
\eqn\abc{
a = - \half \Delta_{12} \, , \qquad b  = \half \Delta_{34} \, , \qquad
c = 0 \, , \qquad \vep = d-2 \, .
}
In general we then seek eigenfunctions of  $D_\vep$, 
$F^{(\vep)}_{\bl_1 \bl_2}(a,b;c;x,z)$, which are symmetric in $x,z$ with 
the boundary behaviour, suppressing unnecessary arguments,
\eqn\Fsim{
F_{\bl_1 \bl_2}(x,z) \limu{z\to 0,x\to 0} x^{\bl_1}z^{\bl_2} \, ,
\quad \bl_1 - \bl_2 = 0,1,2, \dots \, ,
}
where the limit $z\to 0$ is taken first\foot{Since $D_\vep (xz)^p =
(xz)^p ( D_\vep |_{a\to a+p,b \to b+p,c\to c+2p} + 2p(p-1+c -{1\over 2}\vep))$,
and consequently ${(xz)^p F^{(\vep)}_{\bl_1 \bl_2}(a+p,b+p;c+2p;x,z) 
= F^{(\vep)}_{\bl_1{+p}\, \bl_2{+p}}(a,b;c;x,z)}$, the parameter $c$ is
superfluous but it is convenient to keep it here.}. The corresponding 
eigenvalues are then
\eqn\eigenv{
\bl_1 ( \bl_1 + c - 1 ) + \bl_2 ( \bl_2 + c - 1 - \vep ) \, .
}
To match \Glim\
\eqn\lldl{
\bl_1 = \half ( \Delta + \ell) \, , \qquad \bl_2 = \half ( \Delta - \ell) \, ,
}
and then \eigenv\ gives the result for $C_{\Delta,\ell}$ in \eigen.

When $\vep=0$ \defD\ is essentially the sum of two hypergeometric operators
and we may make use of
\eqn\hyper{\eqalign{
x \Big ( x (1-x) {\d^2 \over \d x^2} + \big ( c - (a+b+1)x \big )& 
{\d \over \d x} - ab \Big ) x^p F(p+a,p+b;2p+c;x) \cr 
= {}& p(p+c-1) x^p F(p+a,p+b;2p+c;x) \, . \cr}
}
The eigenfunctions of $D_0$ satisfying \Fsim\ are then
\eqn\eigenz{
x^{\bl_1} z^{\bl _2}  F(\bl_1+a,\bl_1+b;2\bl_1+c;x)
F(\bl_2+a,\bl_2+b;2\bl_2+c;z) + x \leftrightarrow z \, ,
}
with the expected eigenvalue from \eigenv.
With $a,b,c$ as in \abc\ this solution is applicable to
two dimensional conformal field theories where in  Euclidean space
$z$ and $x=z^*$ are the usual complex coordinates.

For $\vep=2$ the solutions are almost as simple making use of
\eqn\Dtt{
D_2 {1\over x-z} = {1\over x-z} \Big ( D_0 \big |_{{a\to a-1, b\to b-1\atop
c\to c-2}} - c + 2 \Big ) \, .
}
Using the known eigenfunctions of $D_0$ the solutions with the  required
leading behaviour for $x,z \sim 0$ are then
\eqn\eigent{
{1\over x-z} \Big ( x^{\bl_1+1} z^{\bl _2}  F(\bl_1+a,\bl_1+b;2\bl_1+c;x)
F(\bl_2+a-1,\bl_2+b-1;2\bl_2+c-2;z) - x \leftrightarrow z \Big ) \, ,
}
with the eigenvalue  given by \eigenv.
The solution given by \eigent\ with \abc\ and $\bl_1,\bl_2$ as in \lldl\
is identical with that in \DO\ for $d=4$.

For $\vep=4$ the results are less simple but they can be derived in a 
similar fashion by using
\eqn\Dts{\eqalign{
& D_4 {1\over (x-z)^3} \cr
&{} = {1\over (x-z)^3} \bigg ( D_0 \big |_{{a\to a-3, b\to b-3\atop
c\to c-6}} - {2xz \over x-z} \Big ( (1-x) {\pr \over \pr x}
- (1-z) {\pr \over \pr z} \Big ) - 3c + 12 \bigg ) \, . \cr}
}
To obtain the solutions in this case we introduce
\eqn\FF{
F^{\pm}_{p,q}(x,z) = x^{p+3}z^{q+3}F(p+a,p+b;2p+c;x) F(q+a,q+b;2q+c;x)
\pm x \leftrightarrow z \, ,
}
and write the eigenfunctions satisfying
\eqn\eigenD{
D_4 F = C F \, ,
}
in the form
\eqn\trial{
F(x,z) = \sum_{p,q} a_{pq} \, {F^-_{pq}(x,z) \over (x-z)^3} \, .
}
Substituting \trial\ in \eigenD\ gives
\eqn\recur{\eqalign{
& \sum_{p,q} a_{pq} \Big ( p(p+c-3)+q(q+c-3)+3c-12-C \Big ) \Big ( {1\over x}
- {1\over z} \Big ) F^-_{pq}(x,z) \cr
&{} = - 2 \sum_{p,q} a_{pq} \Big (
(1-x) {\pr \over \pr x} - (1-z) {\pr \over \pr z} \Big ) F^-_{pq}(x,z) \,.\cr}
}
To solve \recur\ we use
\eqnn\ident
$$\eqalignno{
\!\!\!\!\! \Big ( {1\over x} - {1\over z} \Big ) F^-_{p\, q}(x,z) = {}&
F^+_{p{-1}\,q}(x,z) - F^+_{p\,q{-1}}(x,z) + A_{pq} F^+_{p\, q}(x,z) 
+ B_p F^+_{p{+1}\,q}(x,z) - B_q F^+_{p\,q{+1}}(x,z) \, , \cr
A_{pq}=  {}& (p-q)(p+q+c-1) {\hat A}_{pq} \, , \quad {\hat A}_{pq} =
{2 (2a-c)(2b-c) \over (2p+c)(2p+c-2)(2q+c)(2q+c-2)}\, , \cr
B_p = {}& {(p+a)(p+b)(p+c-a)(p+c-b)\over (2p+c-1)(2p+c)^2(2p+c+1)} \,, &
\ident \cr}
$$
and
\eqn\identd{\eqalign{
\Big (
(1-x) & {\pr \over \pr x} - (1-z) {\pr \over \pr z} \Big ) F^-_{pq}(x,z) \cr
= {}& p \, F^+_{p{-1}\,q}(x,z) - q\, F^+_{p\,q{-1}}(x,z) 
- \half (c-8) A_{pq}\, F^+_{p\, q}(x,z) \cr
&{} - (p+c-4)B_p \, F^+_{p{+1}\,q}(x,z)
+ (q+c-4) B_q \,  F^+_{p\,q{+1}}(x,z) \, , \cr}
}
which follow from identities for ordinary hypergeometric functions,
to obtain recurrence relations for $a_{pq}$ involving $a_{pq},a_{p{\pm 1}q}$
and $a_{pq{\pm 1}}$. For the right value of $C$ this may be solved for
only 5 terms non zero. Choosing a solution with a normalisation determined by
\Fsim\ we have the following non zero terms
\eqnn\resa
$$\eqalignno{
a_{\bl_1 \bl_2{-3}} = {}&  1  \, ,  \qquad
a_{\bl_1{-1} \bl_2{-2}} =  - {\bl_1-\bl_2+3 \over \bl_1-\bl_2+1} \, , \cr
a_{\bl_1{+1} \bl_2{-2}} = {}&
-{\bl_1 + \bl_2 + c -4 \over \bl_1 + \bl_2 + c -2}\, B_{\bl_1}\, , & \resa \cr 
a_{\bl_1 \bl_2{-1}} 
= {}& {(\bl_1 + \bl_2 + c -4)(\bl_1-\bl_2+3)\over
(\bl_1 + \bl_2 + c -2)(\bl_1-\bl_2+1)}\,  B_{\bl_2-2} \, , \cr
a_{\bl_1 \bl_2{-2}} = {}& - (\bl_1 + \bl_2 + c -4)
(\bl_1-\bl_2+3) {\hat A}_{\bl_1 \, \bl_2{-2}}\, , \cr}
$$
with $C$ given by \eigenv\ for $\vep=4$.

Using \lldl\ the result, if $\Delta>2$, for the six-dimensional conformal 
partial wave function satisfying \eigenF\ and \Glim\ is then
\eqnn\Gres
$$\eqalignno{
G_\Delta^{(\ell)} = {}& \F_{00} - {\ell+3 \over \ell+1} \, \F_{{-1}1} \cr
&{} - {\Delta-4\over \Delta-2}\, {(\Delta+\ell-\Delta_{12})
(\Delta+\ell+\Delta_{12})(\Delta+\ell+\Delta_{34})(\Delta+\ell-\Delta_{34})
\over 16 (\Delta+\ell-1)(\Delta+\ell)^2 (\Delta+\ell+1)} \, \F_{11} \cr
&{} + {(\Delta-4)(\ell+3)\over (\Delta-2)(\ell+1)}\cr
&{}\times
{(\Delta-\ell-\Delta_{12}-4)(\Delta-\ell+\Delta_{12}-4)
(\Delta-\ell+\Delta_{34}-4)(\Delta-\ell-\Delta_{34}-4)
\over 16 (\Delta-\ell-5)(\Delta-\ell-4)^2 (\Delta-\ell-3)} \,\F_{02} \cr
&{} + 2(\Delta-4)(\ell+3)\, {\Delta_{12}\,\Delta_{34} \over
(\Delta+\ell) (\Delta+\ell-2) (\Delta-\ell-4)(\Delta-\ell-6)} \, \F_{01} \, , 
& \Gres \cr}
$$
where
\eqnn\Fres
$$\eqalignno{
\F_{nm}(x,z)={}& {(xz)^{{1\over 2}(\Delta-\ell)}\over (x-z)^3} \bigg \{
(-\half x)^\ell \, x^{n+3} z^m \cr
&{}\times F(\half(\Delta+\ell-\Delta_{12})+n,\half(\Delta+\ell+\Delta_{34})+n;
\Delta+\ell+2n;x) \cr
&{}\times 
F(\half(\Delta-\ell-\Delta_{12})-3+m,\half(\Delta-\ell+\Delta_{34})-3+m;
\Delta-\ell-6+2m;z) \cr
&\qquad \qquad \qquad {} - x \leftrightarrow z \bigg \} \, . & \Fres
\cr}
$$
This may perhaps be relevant in the analysis of six dimensional conformal
theories.

\newsec{Solution in terms of Jack Polynomials}

The differential operator $D_\vep$, defined in \defD, is closely related to 
a general class of differential operators acting on symmetric functions of 
several variables, see \refs{\Muir,\Vilenkin}. 
We describe here an approach for finding the eigenfunctions 
of the operator $D_\vep$  which is a modification of the method 
described in \Vilenkin\ and which is potentially applicable
for any $\vep$. To do this we introduce a simpler second order symmetric
operator
\eqn\defDv{
D^J_\vep =  x^2 {\pr^2 \over \pr x^2} +  z^2  {\pr^2 \over \pr z^2}
+ \vep \, {1 \over x-z} \Big ( x^2 {\pr \over \pr x} - z^2
{\pr \over \pr z} \Big ) \, , 
}
so that
\eqn\DD{\eqalign{
D_\vep = {}& - \half \Big [ D^J_\vep ,  x^2 {\pr \over \pr x} + z^2
{\pr \over \pr z} \Big ] + D^J_\vep \cr
&{} + (c-\vep) \Big ( x {\pr \over \pr x} + z {\pr \over \pr z} \Big ) 
- (a+b-\half \vep) \Big ( x^2 {\pr \over \pr x} + z^2 {\pr \over \pr z} \Big ) 
- ab (x+z) \, . \cr}
}
The symmetric eigenfunctions of $D^J_\vep$ satisfy\foot
{$P_{\lambda_1 \lambda_2}(x,z)$ are also eigenfunctions
of the commuting operator $x{\pr\over \pr x} + z{\pr\over \pr z}$,
with eigenvalue $\lambda_1 + \lambda_2$.}, 
\eqn\Deig{
D^J_\vep P_{\lambda_1\lambda_2}(x,z) = \big ( \lambda_1 (\lambda_1-1+\vep)
+ \lambda_2 ( \lambda_2 -1) \big ) P_{\lambda_1\lambda_2}(x,z) \, , \quad
\lambda_1 \ge \lambda_2 \, ,
}
where $P_{\lambda_1 \lambda_2}(x,z)$ is a homogeneous polynomial of degree
 $\lambda_1 + \lambda_2$ in $x,z$, and for $z \to 0$ 
$P_{\lambda_1 \lambda_2}(x,z) \propto x^{\lambda_1}z^{\lambda_2}$.
For $\lambda_1,\lambda_2$ integers,  $P_{\lambda_1\lambda_2}(x,z)$ are just 
the Jack polynomials \Jack\ in two variables. 
It is easy to see from the form of $D^J_\vep$ in \defDv\ that we can take
\eqn\Plc{
(xz)^f  P_{\lambda_1\lambda_2}(x,z) =  P_{\lambda_1{+f}\,\lambda_2{+f}}(x,z)\, ,
}
so that the polynomials $P_{\lambda_1\lambda_2}(x,z)$ may be extended
to arbitrary $\lambda_1,\lambda_2$ with 
$\lambda_1-\lambda_2=0,1,2\dots $. In the two variable case there is
an explicit solution \Vilenkin\ involving Gegenbauer polynomials
\eqn\Psol{
P_{\lambda_1\lambda_2}(x,z) = {\lambda_- ! \over ( \vep)_{\lambda_-}}\,
(xz)^{{1\over 2}(\lambda_1+\lambda_2)}
C^{{1\over 2}\vep}_{\lambda_-}(\si) \, , \quad \lambda_- = \lambda_1 -
\lambda_2 \, , \quad \si = {x+z\over 2(xz)^{1\over 2}} \, , 
}
where we have chosen a somewhat arbitrary normalisation so that
$P_{\lambda_1\lambda_2}(1,1)=1$. In this case
\eqn\Psim{
P_{\lambda_1 \lambda_2}(x,z) \limu{z\to 0} 
{(\half \vep)_{\lambda_-} \over (\vep)_{\lambda_-}} \,
x^{\lambda_1}z^{\lambda_2}  \, .
}

The following are special cases which may be verified directly (note
$D^J_2{1\over x-z} = {1\over x-z}D^J_0$),
\eqn\Pss{\eqalign{
P_{\lambda_1\lambda_2}(x,z)\big |_{\vep=0} = {}&  \half \big (
x^{\lambda_1} z^{\lambda_2} + x^{\lambda_2} z^{\lambda_1} \big )  \, , \cr
P_{\lambda_1\lambda_2}(x,z)\big |_{\vep=2} ={}& {1\over \lambda_- \! +1}
\bigg ({ x^{\lambda_1 +1 } z^{\lambda_2} - x^{\lambda_2} z^{\lambda_1 +1} \over
x-z}\bigg )  = {(xz)^{\lambda_2} \over \lambda_- \! +1}
\bigg ({ x^{\lambda_- +1 } - z^{\lambda_- +1} \over x-z}\bigg ) \, , \cr
P_{\lambda_1\lambda_2}(x,z)\big |_{\vep=4} ={}& {6\over \lambda_- \! + 2}
\bigg (  
{ x^{\lambda_1 + 3 } z^{\lambda_2} - x^{\lambda_2} z^{\lambda_1 +3} \over
( \lambda_- \! + 3)(x-z)^3 } - 
{ x^{\lambda_1 + 2 } z^{\lambda_2+1 } - x^{\lambda_2 +1 } z^{\lambda_1 +2} \over
(\lambda_- \! + 1)(x-z)^3}\bigg )   .  \cr}
}

The Jack polynomials play the role of an extension of simple powers of $x$ for
single variable functions when 
the series expansion of multi-variable symmetric functions is considered.
For subsequent use there are important recurrence relations which may be
verified by using standard identities \Grad\ for 
$\si C^{{1\over 2}\vep}_n ( \si)$ and 
$(1-\si^2)  C^{{1\over 2}\vep}_n {}' (\si)$ as well as its
defining differential equation,
\eqn\recurC{\eqalign{
(x+z)P_{\lambda_1\lambda_2}(x,z) = {}& 
{\lambda_-\! + \vep \over \lambda_-\! + \half \vep} \, 
P_{\lambda_1{+1}\,\lambda_2}(x,z) + {\lambda_- \over \lambda_-\! + \half \vep}
\, P_{\lambda_1\, \lambda_2{+1}}(x,z) \, , \cr
\Big ( x^2 {\pr \over \pr x} + z^2 {\pr \over \pr z} \Big )
P_{\lambda_1\lambda_2}(x,z) = {}& 
{\lambda_1 (\lambda_-\! + \vep )\over \lambda_-\! + \half \vep}\,
P_{\lambda_1{+1}\,\lambda_2}(x,z) + { (\lambda_2 - \half \vep ) 
\lambda_- \over \lambda_-\! + \half \vep}
\, P_{\lambda_1\, \lambda_2{+1}}(x,z) \, . \cr}
}
These are valid for any $\lambda_1-\lambda_2 = 0,1,2,\dots$, for $\lambda_1
=\lambda_2$ the second term on the right hand side is absent 
($P_{\lambda\lambda}(x,z)=(xz)^{\lambda},\,
P_{\lambda{+1}\,\lambda}(x,z)=(xz)^{\lambda}\half(x+z)$).
Using \recurC\ with \DD\ and \Deig\ we may find,
\eqn\DP{\eqalign{
D_\vep P_{\lambda_1\lambda_2} = &{} \big ( \lambda_1(\lambda_1+c-1) +
 \lambda_2(\lambda_2+c-1 - \vep ) \big ) P_{\lambda_1\lambda_2} \cr
&{}- {\lambda_- +\vep \over \lambda_- + \half \vep} \,
(\lambda_-+a)(\lambda_-+b)\,P_{\lambda_1{+1}\,\lambda_2} \cr
&{} - {\lambda_- \over \lambda_- + \half \vep} \,
(\lambda_- -\half\vep+a)(\lambda_- -\half\vep+b)\,P_{\lambda_1\,\lambda_2{+1}}
\, . \cr}
}

We then seek a solution for the eigenfunction for $D_\vep$, with the
limiting behaviour given by \Fsim, in the form
\eqn\Feig{
F_{\lambda_1\lambda_2} = \sum_{m,n\ge 0} r_{mn} \, 
P_{\lambda_1{+m}\, \lambda_2{+n}} \, .
}
The necessary recurrence relation is simplified by taking
\eqn\rrh{
r_{mn} = (\lambda_1+a)_m (\lambda_1+b)_m
(\lambda_2-\half\vep+a)_n (\lambda_2-\half\vep+b)_n \, \hr_{mn} \, ,
}
and then, with the eigenvalue given by \eigenv, we obtain 
\eqn\recurr{\eqalign{
\big ( m & (2\lambda_1 + c-1+m) + n(2\lambda_2 + c-1 - \vep +n )\big
)\hr_{mn}\cr 
&{} = {\lambda_- +m-n - 1 + \vep \over \lambda_- +m-n - 1 + \half
\vep} \, \hr_{m{-1}\, n} 
+ {\lambda_- +m-n + 1 \over \lambda_- + m-n +  1 + \half \vep} \,
\hr_{m\,n{-1}}\, . \cr} }
This may be solved iteratively starting from $\hr_{00}$. The general
solution is rather involved so we consider first some special cases.
It is easy to see that
\eqn\rzero{
\hr_{m0} = {(\vep)_{\lambda_-\! + m} \over m! (2 \lambda_1 +c)_m 
(\half \vep)_{\lambda_-\! + m}} \, ,
}
choosing $\hr_{00}$ so that from \Psim\ we have
\eqn\Fsim{
F_{\lambda_1 \lambda_2}(x,z) \limu{z\to 0}
x^{\lambda_1}z^{\lambda_2} \, F(\lambda_1+a, \lambda_1+b ;
2 \lambda_1+c ; x ) \, .
}

For $\vep=0,2$ respectively the solutions of \recurr\ are quite simple
\eqn\solr{
\hr_{mn} = {2\over m!\,n!(2\lambda_1+c)_m(2\lambda_2+c)_n} \, , \ \
\hr_{mn} = 
{\lambda_-+1+m-n\over m!\,n!(2\lambda_1+c)_m(2\lambda_2+c-2)_n} \, , 
}
where we match \rzero\ for $n=0$.
Substituting in \Feig\ and using the relevant results from \Pss\ we easily
recover the solutions \eigenz\ and \eigent. For $\vep=4$ the result is
\eqn\solfour{\eqalign{
\hr_{mn} = {}& {1\over m!\,n!(2\lambda_1+c)_m(2\lambda_2+c-4)_n} \,
{\lambda_-+2+m-n\over (\lambda_1+\lambda_2+c-2)(\lambda_- +1)}\cr
&{}\times {\ts{1\over 6}}\Big ( (\lambda_1+\lambda_2+c-2)\big ( (\lambda_- +1) 
(\lambda_- + 3+ m-n ) - 2n \big ) - 2mn \Big ) \, . \cr}
}
With the corresponding result for $P_{\lambda_1\lambda_2}$ from \Pss\
we may also obtain the solution given by \trial\ and \resa\ if we
recognise that $\hr_{m{-1}\, n} - \hr_{m\,n{-1}}$ contains a factor
$\lambda_- + m - n + 2$ so that the summand is expressible as a linear
combination of five expressions in which the dependence on $m,n$ factorises
in an appropriate fashion.

In the above results for $\vep=0,2,4$ the summations over $m,n$ are 
unrestricted so that they involve $P_{\lambda_1\lambda_2}$ for 
$\lambda_1 < \lambda_2$. However for $\vep$ even we may note the symmetry
\eqn\symP{
P_{\lambda_1\, \lambda_2} = 
P_{\lambda_2{-{1\over 2}\vep}\,\lambda_1{+{1\over 2}\vep}} \, .
}
In consequence in \Feig\ we may take instead of the results given by
\rrh\ and \solr\ or \solfour,
\eqn\rnew{\eqalign{
\hr_{mn}^{\vphantom g} \to {}& \hr_{mn}^{\vphantom g} + 
\hr_{n{-\lambda_-}{-{1\over 2}\vep}\, m {+\lambda_-}{+{1\over 2}\vep}} \, ,
\qquad \vep > 0 \, , \cr
\hr_{mn}^{\vphantom g} \to {}& \cases{\hr_{mn}^{\vphantom g} +
\hr_{n{-\lambda_-}\, m {+\lambda_-}}\, , & $\lambda_-+m-n > 0 \, ,$ \cr
\hr_{mn}^{\vphantom g}\, , & $\lambda_-+m-n = 0 \, ,$ \cr} \qquad \vep  =0
\, , \cr}
}
with the summation now restricted to $\lambda_-+m-n\ge 0$. Except for $\vep=0$
we should note that $\hr_{mn}=0$ if $\lambda_-+m-n + \half \vep =0$ and
contributions for $-\vep < \lambda_- + m-n < 0$, for which both contributions
in \rnew\ do not satisfy $\lambda_-+m-n\ge 0$, cancel. When $\vep=2$ the
resulting expansion is in terms of Schur polynomials as considered by 
Heslop and Howe \Howe\ for the four-dimensional operator product expansion.

In general the recurrence relations \recurr\ can  be solved in the form
\eqnn\solrh
$$\eqalignno{
\hr_{mn} = {}& 
{(\lambda_-\!+ m)! \, (\vep)_{\lambda_-\!+ m-n} \over  m!\, n! \, 
(\lambda_- \! + m - n)! \, (2\lambda_1 + c )_{m}
(2\lambda_2 + c - \half \vep )_{n} (\half \vep)_{\lambda_- \! + m+1}} \cr
&{} \times  ( \lambda_- \! + m -n + \half \vep ) \,
{(-\lambda_1 - \lambda_2 -c +1 + \half \vep )_{\lambda_-}(\vep)_{\lambda_-}
\over (- \lambda_1 - \lambda_2 -c +1 +  \vep )_{\lambda_-}
(\half\vep)_{\lambda_-}}  \cr
& \qquad \quad{}\times {}_4 F_3 \bigg ( { - \lambda_- , - \lambda_- \! -  m +n ,
\lambda_1 +\lambda_2 +c -1 , \half \vep \atop
- \lambda_- \! - m , 2\lambda_2 +c - \half \vep + n ,\, \vep } ; 1 \bigg )\,,
& \solrh \cr}
$$
which matches \rzero\ for $n=0$. 
The terminating ${}_4 F_3$ function in \solrh\ has $\lambda_- \! +1$
terms in its expansion, explicit straightforward results may be obtained for
$\lambda_- = 0,1$. The solution \solrh\ of the recurrence relation \recurr\ 
was obtained initially by adapting the results of Koornwinder and 
Sprinkhuizen-Kuyper \Koo, but an independent proof is given in appendix A.

\newsec{Compact Case}

The above discussion is an extension of known results for two variable
orthogonal polynomials. These also arise \nirschl\ in the expansion of four
point correlation functions when the fields belong to tensor representations
of $O(n)$. We here describe some analogous results to make the comparison
clear. The relevant differential operation representing the Casimir 
operator, or Laplacian, acting on symmetric functions of two variables 
has the form,
\eqnn\defDc
$$\eqalignno{
{\tilde D}_\vep = {}& 
- x (1-x) {\pr^2 \over \pr x^2} -  z (1-z) {\pr^2 \over \pr z^2}
- (a+1) \Big ( {\pr \over \pr x} +  {\pr \over \pr z} \Big )
+ (a+b+2) \Big ( x {\pr \over \pr x} + z {\pr \over \pr z} \Big ) \cr
&{} - \vep \, {1 \over x-z} \Big ( x(1-x) {\pr \over \pr x} - z(1- z)
{\pr \over \pr z} \Big ) + \half ( a+b)(a+b+\vep + 2) \, , & \defDc \cr}
$$
for $a,b$ integers and $\vep=n-4$. This may be be written in the form
\eqn\Dad{\eqalign{
{\tilde D}_\vep = {}& - {1\over w} \Big ( {\pr \over \pr x} w\, x(1-x) 
{\pr \over \pr x} + {\pr \over \pr z} w\, z(1-z)  {\pr \over \pr z}
\Big ) + \half ( a+b)(a+b+\vep + 2) \, , \cr 
&{} w = x^a z^a (1-x)^b(1-z)^b (x-z)^\vep \, , \cr}
}
so that the symmetric eigenfunctions are orthogonal, if $\vep>0$, integrated 
over $0<z < x < 1$. For $\vep=2,4$ solutions were obtained
by Vretare \Vretare\ in terms of Jacobi polynomials which are similar to
\eigent\ and \trial\ with \resa.

For general $\vep$ we may obtain and expansion in terms of Jack
polynomials as in section 3. Instead of \DD\ we may now write
\eqn\DDc{\eqalign{
{\tilde D}_\vep = {}& \half \Big [ D^J_\vep ,  {\pr \over \pr x} + 
{\pr \over \pr z} \Big ] + D^J_\vep \cr
&{} - (a+1) \Big (  {\pr \over \pr x} +  {\pr \over \pr z} \Big )
+ (a+b + 2) \Big (x {\pr \over \pr x} + z {\pr \over \pr z} \Big )
+ \half ( a+b)(a+b+\vep + 2)  \, . \cr}
}
The relevant equation for Jack polynomials is then
\eqn\dJ{
\Big ( {\pr \over \pr x} + {\pr \over \pr z} \Big )
P_{\lambda_1\lambda_2}(x,z) = 
{(\lambda_1 + \half \vep) \lambda_- \over \lambda_-\! + \half \vep}\,
P_{\lambda_1{-1}\,\lambda_2}(x,z) + { \lambda_2 
(\lambda_- +\vep ) \over \lambda_-\! + \half \vep}
\, P_{\lambda_1\, \lambda_2{-1}}(x,z) \,  .
}

The polynomial eigenfunctions have the form
\eqn\eigenc{
{\tilde F}_{\lambda_1\lambda_2} = 
\sum_{m=0}^{\lambda_1} \sum_{n=0,n\le m}^{\lambda_2}\!\! s_{mn} \, P_{mn} \, ,
}
and for the series to truncate the eigenvalue of \DDc\ must be
\eqn\eigenD{
\big ( \lambda_1 + \half (a+b) \big ) \big ( \lambda_1  + \half (a+b) 
+ 1 + \vep \big ) + \big ( \lambda_2 + \half (a+b) \big ) \big (
\lambda_2  + \half (a+b) + 1 \big ) \, .
}
If we write
\eqn\rrc{
s_{mn} = {1\over (1+ \half \vep)_m n!\,  (a+1 + \half \vep )_m 
(a+1)_n } \, \hs_{mn} \, ,
}
the necessary recurrence relation simplifies to
\eqn\recurc{\eqalign{
\big ( (\lambda_1 -m)& (\lambda_1 + a+b+1 + \vep + m) + (\lambda_2 -n)
(\lambda_2 + a+b + 1 +n )\big
)\hs_{mn}\cr
&{} = - { m-n - 1 + \vep \over m-n - 1 + \half \vep} \, \hs_{m\, n{+1}}
- {m-n + 1 \over m-n + 1 + \half \vep} \, \hs_{m{+1} \,n}\, . \cr }
}
In this case we may solve for $\hs_{mn}$ starting just from 
$\hs_{\lambda_1\lambda_2}$. The recurrence relation \recurc\ is 
related to \recurr\ by $m\to \lambda_2-n , \, n\to \lambda_1-m , \, 
\lambda_1 \leftrightarrow -  \lambda_2, \, c \to - (a+b)$ so that 
we can then use the same solution with $\hs_{mn} \propto \hr_{\lambda_2{-n}\,
\lambda_1{-m}}$ (also in \rrh\ and \rrc\ we may let $a\to -a , \, b\to 0$). 
This gives
\eqnn\solsh
$$\eqalignno{
\hs_{m n} = {}& (-1)^{m+n} 
{ (\lambda_1 -n )! \, (\lambda_1 + a+b +1 +\half \vep)_{m}
(\lambda_2 + a+b +1 )_{n} (\vep)_{m-n} \over (\lambda_1 - m)! \, 
(\lambda_2 - n)!\, (m-n)! \, (\half \vep)_{\lambda_1 - n+1}} \cr
&{} \times  (m - n + \half \vep ) \,  (\half \vep)_{\lambda_1} \lambda_2!\cr
& \qquad \quad{}\times {}_4 F_3 \bigg ( { - \lambda_- , n- m ,
- \lambda_1 - \lambda_2 - a -b -1 , \half \vep \atop 
- \lambda_1 + n , -\lambda_1 -a-b - \half \vep - m ,\, \vep }; 1 \bigg )\, , 
& \solsh \cr}
$$
where we have chosen a normalisation such that $\hs_{00}=1$.

\newsec{Recurrence Relations for Conformal Amplitudes}

With the aid of the solution obtained in section 3 we may determine
recurrence relations for conformal amplitudes which are valid for
general dimension. We first exhibit
\eqn\recurF{\eqalign{
\!\!\!\!\!& \Big ( {x+z\over xz} - 1 \Big ) F_{\lambda_1\lambda_2}(x,z) = 
F_{\lambda_1\lambda_2{-1}}(x,z) + {\lambda_-(\lambda_- \! -1 + \vep)
\over (\lambda_- \!  + \half \vep ) (\lambda_- \!  -1 + \half \vep)}
F_{\lambda_1{-1}\lambda_2}(x,z) \cr
\!\!\!\!\!&{}- {\ts{1\over 4}}
\big ( (2\lambda_1+c)(2\lambda_1+c-2)+(2\lambda_2+c - \vep)
(2\lambda_2+c - 2 -\vep ) \cr
\noalign{\vskip -3pt}
\!\!\!\!\! & \hskip 8cm {} - \vep(\vep -2) \big ) 
{\hat A}_{\lambda_1\lambda_2{-{1\over 2}\vep}}\, F_{\lambda_1\lambda_2}(x,z) \cr
\!\!\!\!\! &{}+{ (\lambda_1+\lambda_2+c-1)(\lambda_1+\lambda_2+c-\vep)\over
(\lambda_1+\lambda_2+c- \half \vep )(\lambda_1+\lambda_2+c-1-\half \vep)}\,
B_{\lambda_1}  F_{\lambda_1{+1}\lambda_2}(x,z) \cr
\!\!\!\!\! &{}+ {\lambda_-(\lambda_- \! -1 + \vep)
\over (\lambda_- \!  + \half \vep ) (\lambda_- \!  -1 + \half \vep)}\,
{ (\lambda_1+\lambda_2+c-1)(\lambda_1+\lambda_2+c-\vep)\over
(\lambda_1+\lambda_2+c- \half \vep )(\lambda_1+\lambda_2+c-1-\half \vep)}\,
B_{\lambda_2 - {1\over 2}\vep} F_{\lambda_1\lambda_2{+1}}(x,z) \, ,
\cr}
}
with $B_p$ and ${\hat A}_{pq}$ as in  \resa.
This result reduces to the special case for $(1-v)G_\Delta^{(\ell)}(u,v)/u$
given in  \SCFT\ when $\vep=2$.

More intricately we have
\eqn\recurFF{\eqalign{
& \Big ( {1\over xz} - {x+z\over 2\, xz} + {1\over 4} \Big ) 
F_{\lambda_1\lambda_2}(x,z) \cr
&{} =
F_{\lambda_1{-1}\lambda_2{-1}}(x,z) - 
A_{\lambda_1} \, F_{\lambda_1\lambda_2{-1}}(x,z)
- {\lambda_-(\lambda_- \! -1 + \vep)
\over (\lambda_- \!  + \half \vep ) (\lambda_- \!  -1 + \half \vep)}
A_{\lambda_2- {1\over 2}\vep} \, F_{\lambda_1{-1}\lambda_2}(x,z) \cr
&\quad {}+ B_{\lambda_1} \, F_{\lambda_1{+1}\lambda_2{-1}}(x,z)
+ {\lambda_-(\lambda_- \! -1) (\lambda_- \! -1 + \vep)(\lambda_- \! -2 + \vep)
\over (\lambda_- \!  + \half \vep ) (\lambda_- \!  -1 + \half \vep)^2
(\lambda_- \!  -2 + \half \vep) }
B_{\lambda_2 - {1\over 2} \vep } \, F_{\lambda_1{-1}\lambda_2{+1}}(x,z) \cr
& \quad {} + \bigg ( A_{\lambda_1}^{\vphantom a}
A_{\lambda_2- {1\over 2}\vep} \cr
& \quad {} - {\vep(\vep-2)\over 32(\lambda_- \! -1 + \half \vep)
(\lambda_- \! + 1 + \half\vep)(\lambda_1+\lambda_2 + c -\half\vep)
(\lambda_1+\lambda_2 + c -2 -\half\vep)}\cr
& \quad {}\times 
\Big ( \half (2\lambda_1+c)(2\lambda_1+c-2)+ \half (2\lambda_2+c - \vep) 
(2\lambda_2+c - 2 -\vep ) - (2a-c)^2 - (2b-c)^2 \cr
\noalign{\vskip -4pt}
& \hskip 3cm {}  + (2a-c)(2b-c)( A_{\lambda_1}^{\vphantom a} +
A_{\lambda_2- {1\over 2}\vep}) -16\, A_{\lambda_1}^{\vphantom a} 
A_{\lambda_2- {1\over 2}\vep} \Big ) \bigg )F_{\lambda_1\lambda_2}(x,z)  \cr
&\quad {} - {(\lambda_1+\lambda_2 + c -1)
(\lambda_1+\lambda_2 + c -\vep)\over (\lambda_1+\lambda_2 + c -\half\vep)
(\lambda_1+\lambda_2 + c -1 - \half \vep)}\bigg (  B_{\lambda_1}^{\vphantom a} 
A_{\lambda_2- {1\over 2}\vep}  \, F_{\lambda_1{+1}\lambda_2}(x,z) \cr
\noalign{\vskip -3pt}
&\hskip 5cm  {} + {\lambda_-(\lambda_- \! -1 + \vep)
\over (\lambda_- \!  + \half \vep ) (\lambda_- \!  -1 + \half \vep)}
B_{\lambda_2 - {1\over 2}\vep} 
A_{\lambda_1}^{\vphantom a}  \, F_{\lambda_1\lambda_2{+1}}(x,z) \bigg ) \cr
&\quad {} + {(\lambda_1+\lambda_2 + c)(\lambda_1+\lambda_2 + c-1)
(\lambda_1+\lambda_2 + c +1 -\vep)
(\lambda_1+\lambda_2 + c -\vep)\over (\lambda_1+\lambda_2 + c -1 - \half \vep)
(\lambda_1+\lambda_2 + c -\half\vep)^2
(\lambda_1+\lambda_2 + c + 1 - \half \vep)} \cr
\noalign{\vskip -3pt}
&\hskip 8cm  {} \times B_{\lambda_1}^{\vphantom a} 
B_{\lambda_2 - {1\over 2}\vep}  \, F_{\lambda_1{+1}\lambda_2{+1}}(x,z)\, ,
\cr}
}
where we here define
\eqn\defAp{
A_p = {(2a-c)(2b-c)\over 2(2p+c)(2p+c-2)} \, .
}
Again this result reduces to the special case for 
$(1+v)G_\Delta^{(\ell)}(u,v)/u$ given in  \SCFT\ when $\vep=2$.

Both \recurF\ and \recurFF\ can be proved using \Plc\ and the first 
result in \recurC\ in the expansion in terms of Jack polynomials
together with various results for the
expansion coefficients which are listed in the appendix.

The formula obtained for the solution when $\vep=4$ is also a special case
of a relation between the results for $\vep$ and $\vep+2$ which takes
the form, for a normalisation given by \Fsim,
\eqnn\relep\hskip -1cm
{$$\eqalignno{
{(x-z)^2 \over (xz)^2} F_{\lambda_1\lambda_2}^{(\vep{+2})} & (x,z) 
= F_{\lambda_1\,\lambda_2{-2}}^{(\vep)}(x,z) -
{(\lambda_- \! + \vep)(\lambda_- \! + 1 +  \vep) \over 
(\lambda_- \! + \half \vep )( \lambda_- \! + 1 + \half \vep)}
F_{\lambda_1{-1}\,\lambda_2{-1}}^{(\vep)}(x,z) \cr
&\quad {} - {(\lambda_1+\lambda_2 + c -1 -\vep)
(\lambda_1+\lambda_2 + c -2 -\vep)\over (\lambda_1+\lambda_2 + c -1 -\half\vep)
(\lambda_1+\lambda_2 + c -2 - \half \vep)} & \relep \cr
& \qquad {}\times \bigg ( B_{\lambda_1}^{\vphantom a} 
F_{\lambda_1{+1}\,\lambda_2{-1}}^{(\vep)}(x,z) - 
{(\lambda_- \! + \vep)(\lambda_- \! + 1 +  \vep) \over 
(\lambda_- \! + \half \vep )( \lambda_- \! + 1 + \half \vep)}
B_{\lambda_2{-1}{-{1\over 2}\vep}} \,
F_{\lambda_1\,\lambda_2}^{(\vep)}(x,z) \bigg ) \cr
&\quad {} - (\lambda_1+\lambda_2 + c -2 -\vep) (\lambda_- \! + 1+ \vep )
{\hat A}_{\lambda_1\, \lambda_2{-1}{-{1\over 2}\vep}} \,  
F_{\lambda_1\,\lambda_2{-1}}^{(\vep)}(x,z) \, . \cr} 
$$}
This may be obtained from
an analogous relation for $P_{\lambda_1\lambda_2}$,
\eqn\recurP{
{(x-z)^2 \over (xz)^2} P_{\lambda_1\lambda_2}^{(\vep{+2})}(x,z) =
{2(1+\vep) \over \lambda_- \! + 1 + \half \vep} \Big (
P_{\lambda_1\,\lambda_2{-2}}^{(\vep)}(x,z) - 
P_{\lambda_1{-1}\,\lambda_2{-1}}^{(\vep)}(x,z) \Big ) \, ,
}
which follows from standard identities for $C_n^{{1\over 2}\vep}(\si)$.

\newsec{Conclusion}

We have endeavoured to obtain detailed expressions for conformal 
partial wave amplitudes which may be relevant in the operator product
expansion analysis of conformal four point functions. The results for
two and four dimensions, which confirm those obtained earlier in \DO,
are as simple as may reasonably be expected. For six dimensions the
results obtained here do not have any apparent further simplifications,
except in possible special cases. However for general dimensions, in 
particular for three,
the results given as an expansion in terms of Jack polynomials, while
related to similar results in the mathematical literature, still leave
some things to be desired which might be expected to hold in  more explicit 
expressions. In particular we may note the symmetry properties of the four 
point function \fourp\ under interchange of fields which is reflected in 
corresponding properties of the conformal amplitudes. To exhibit these
we define
\eqn\defxzn{
x' = {x \over x-1} \, , \qquad z' = {z \over z-1} \, ,
}
and then, from the form of the operator in \defD, we have
\eqn\Dxz{
\big ( (1-x)(1-z) \big )^{-b} D_{\vep,x',z'} \big ( (1-x)(1-z) \big )^{b} 
= D_{\vep,x,z} \big |_{a\to c-a} \, ,
}
and also correspondingly for $a\leftrightarrow b$. Hence we have
\eqn\FFs{
(-1)^{\lambda_1+\lambda_2} F^{(\vep)}_{\lambda_1\lambda_2}(a,b;c;x',z') = 
\big ( (1-x)(1-z) \big )^{b} F^{(\vep)}_{\lambda_1\lambda_2}(c-a,b;c;x,z) \, ,
}
which corresponds in \fourp\ to taking $\phi_1(\eta_1) \leftrightarrow
\phi_2(\eta_2)$. The analogous result obtained when $a\leftrightarrow b$
is also interpretated as corresponding to  $\phi_3(\eta_3) \leftrightarrow
\phi_4(\eta_4)$. This symmetry is easily demonstrated in the explicit 
solutions for $d=2,4,6$ but is necessarily hidden in the Jack polynomial
expansion. 

On the other hand we may note that Jack polynomials have a ready extension
to superspace \SJack\ so that there should be corresponding version of 
superconformal partial wave amplitudes.

\bigskip
\noindent
{\bf Acknowledgements}
\medskip

We are grateful to Professor Koornwinder for sending us a copy of ref.
\Koo\ so as to allow us to find the solution \solrh. We are also happy
to thank Emery Sokatchev for stimulating discussions.

\appendix{A}{Solution of Recurrence Relation}

The ${}_4 F_3$ function appearing in \solrh\ is balanced or Saalsch\"utzian,
i.e. we have
\eqn\bal{
{}_4 F_3 \Big ( { {a,b,c,d \atop e,f,g}}; 1 \Big ) \, , \qquad
a+b+c+d+1 = e+f+g \, ,
}
with one of $a,b,c,d$ a negative integer. In this case there are important
contiguous relations, see section 3.7 in \Spec. The crucial one has
the form
\eqn\contig{\eqalign{
& e(f-1)(e-a)(g-a) \bigg ( 
{}_4 F_3 \Big ( {a{-1},b,c,d \atop e,f{-1},g}; 1 \Big )
- {}_4 F_3 \Big ( {a,b,c,d \atop e,f,g}; 1 \Big ) \bigg )\cr
&{} + a(e-b)(e-c)(e-d) \bigg ( 
{}_4 F_3 \Big ( {a{+1},b,c,d \atop e{+1},f,g}; 1 \Big )
- {}_4 F_3 \Big ( {a,b,c,d \atop e,f,g}; 1 \Big ) \bigg ) \cr
&\quad {} = - (e-a)bcd\, 
{}_4 F_3 \Big ( {a,b,c,d \atop e,f,g}; 1 \Big ) \, . \cr}
}
If we define
\eqn\Fmn{
F_{mn} = {}_4 F_3 
\bigg ({-\lambda_- , -\lambda_- \! - m +n,\lambda_1 +\lambda_2 +c -1 ,
\half \vep \atop
- \lambda_- \! - m , 2\lambda_2 +c - \half \vep + n ,\, \vep } ; 1 \bigg )\,,
}
then \contig\ gives
\eqn\Frecur{\eqalign{
& n(\lambda_- \! +m)(\lambda_- \! +m -n +\vep )(2\lambda_2 +c-1-\half \vep + n)
F_{m\,n{-1}} \cr
&{} + m(\lambda_- \! +m + \half \vep )(\lambda_- \! +m -n )
(2\lambda_1 +c-1 + m) F_{m{-1}\,n} \cr
&{} = (\lambda_- \! +m) (\lambda_- \! +m -n + \half \vep )
\big ( m (2\lambda_1 + c-1+m) + n(2\lambda_2 + c-1 - \vep +n )\big ) F_{mn} \,.
\cr}
}
It is then straightforward to verify that \solrh\ satisfies \recurr.
It is evident that $F_{m0}$ reduces to a ${}_3F_2$ function and
we have
\eqn\Fzero{
F_{m0}= {(-\lambda_1 - \lambda_2 -c +1 + \vep )_{\lambda_-}
(\half \vep)_{\lambda_-}
\over (- \lambda_1 - \lambda_2 -c +1 + \half \vep )_{\lambda_-}
(\vep)_{\lambda_-}} \, .
}

We may also use contiguous relations to obtain formulae involving
changes in $\lambda_1, \lambda_2$ by $\pm 1$ and also $\vep$ by 2.
These were used to prove the various relations given in section 5
but are rather long winded to write down. We here present them for
$\br^{(\vep)}_{\lambda_1\lambda_2,mn}$ which is a solution of \recurr\ with 
$\br^{(\vep)}_{\lambda_1\lambda_2,00}=1$, $\hr_{mn} = (\vep)_{\lambda_-}
\br^{(\vep)}_{\lambda_1\lambda_2,mn}/(\half \vep)_{\lambda_-}$,
\eqnn\rrr
$$\eqalignno{
&{} (m+1) (\lambda_- \! - n- 1 + \half \vep)\, 
\br^{(\vep)}_{\lambda_1{-1}\lambda_2,m{+1}n} \cr
&{} + (n+1) (\lambda_- \! +m + 1 + \half \vep)\, 
\br^{(\vep)}_{\lambda_1\lambda_2{-1},mn{+1}} \cr
&{} - 2 (\lambda_- \! + \half \vep)(\lambda_1 +\lambda_2 + c -1 -\vep) \cr
\noalign{\vskip -2pt}
& \hskip 2cm {}\times {(2\lambda_1 +2m +c) (2\lambda_2 +2n +c -\vep) \over
(2\lambda_1 +c -2)(2\lambda_1 +c) (2\lambda_2 +c -2 -\vep)
(2\lambda_2 +c -\vep)} \, \br^{(\vep)}_{\lambda_1\lambda_2,mn} \cr
&{} + { (\lambda_1 +\lambda_2 + c -1 -\vep) (\lambda_1 +\lambda_2 + c -\vep) 
\over (\lambda_1 +\lambda_2 + c -1 - \half\vep)
(\lambda_1 +\lambda_2 + c - \half\vep)} 
(\lambda_1 +\lambda_2 + c +m - \half \vep)\cr
\noalign{\vskip -2pt}
& \hskip 2cm {}\times { 2\lambda_2 + n +c -\vep \over
(2\lambda_2 +c -1 -\vep)(2\lambda_2 +c -\vep)^2
(2\lambda_2 +c +1 -\vep)} \, \br^{(\vep)}_{\lambda_1\lambda_2{+1},mn{-1}} \cr
&{} - { (\lambda_1 +\lambda_2 + c -1 -\vep) (\lambda_1 +\lambda_2 + c -\vep)
\over (\lambda_1 +\lambda_2 + c -1 - \half\vep)
(\lambda_1 +\lambda_2 + c - \half\vep)} 
(\lambda_1 +\lambda_2 + c + n - \half \vep)\cr
\noalign{\vskip -2pt}
& \hskip 2cm {}\times { 2\lambda_1 + m +c \over
(2\lambda_1 +c -1 )(2\lambda_1 +c )^2
(2\lambda_1 +c +1 )} \, \br^{(\vep)}_{\lambda_1{+1}\lambda_2,m{-1}n} =0 \, . 
& \rrr \cr }
$$
For a change in $\vep$,
\eqnn\rra
$$\eqalignno{
&{} (m+1) (2\lambda_1+c-2)\,
\br^{(\vep)}_{\lambda_1{-1}\lambda_2,m{+1}n} \cr
&{} - (\lambda_- \! +m + 1 + \half \vep)(2\lambda_2+c-2 -\vep) \,
\br^{(\vep)}_{\lambda_1\lambda_2{-1},mn{+1}} \cr
&{} + { (\lambda_1 +\lambda_2 + c -1 -\vep) (\lambda_1 +\lambda_2 + c -\vep)
\over (\lambda_1 +\lambda_2 + c -1 - \half\vep)
(\lambda_1 +\lambda_2 + c - \half\vep)}
(\lambda_1 +\lambda_2 + c +m - \half \vep) \cr
\noalign{\vskip -2pt}
& \hskip 2cm {}\times { 1 \over
(2\lambda_2 +c - 1 -\vep)(2\lambda_2 + c - \vep)(2\lambda_2 +c +1 -\vep)} \, 
\br^{(\vep)}_{\lambda_1\lambda_2{+1},mn{-1}} \cr
&{} - {(\lambda_1 +\lambda_2 + c -1 -\vep) (\lambda_1 +\lambda_2 + c -\vep) 
\over (\lambda_1 +\lambda_2 + c -1 - \half\vep)
(\lambda_1 +\lambda_2 + c - \half\vep)} \cr
\noalign{\vskip -2pt}
& \hskip 2cm {}\times { 2\lambda_1 + m +c \over
(2\lambda_1 +c -1 )(2\lambda_1 +c )
(2\lambda_1 +c +1 )} \, \br^{(\vep)}_{\lambda_1{+1}\lambda_2,m{-1}n} \cr
&{}+ (\lambda_- \! + \half \vep) { \lambda_- + m-n + \half \vep \over
\lambda_- + m-n -1 + \half \vep} (2\lambda_2 +2n +c -\vep)  \, 
\br^{(\vep+2)}_{\lambda_1\lambda_2{+1},mn{+1}} = 0 \, . & \rra \cr} 
$$
There is also an associated independent relation obtained from \rra\
by  ${\lambda_1 \to \lambda_2 - \half \vep}$, ${\lambda_2 \to \lambda_1 + 
\half \vep}$, $ m \leftrightarrow n $ and then letting 
$\br^{(\vep)}_{\lambda_2{-{1\over 2}} \vep \, \lambda_1{+{1\over 2}} \vep,
nm} \to \br^{(\vep)}_{\lambda_1\lambda_2,mn}$ (\rrr\ is unchanged under
this transformation).

\listrefs
\bye